\documentclass[12pt]{article}
\usepackage{lineno}%
{\par\runninglinenumbers
\usepackage{graphicx}
\usepackage[top=3cm, bottom=2cm, left=3cm, right=3cm]{geometry}

\title{A review of the relevance of the `CLOUD' results and other recent observations
 to the possible effect of cosmic rays on the terrestrial climate.}
\author{A. D. Erlykin $^{1,3}$, T. Sloan $^2$ and A. W. Wolfendale $^{3}$\\
$(1)$ P.N.Lebedev Physical Institute, Leninsky prosp., Moscow, Russia\\
corresponding author: tel.+74991358737, e-mail: erlykin@sci.lebedev.ru\\
$(2)$ Physics Department, University of Lancaster, Lancaster, UK\\
$(3)$ Department of Physics, Durham University, Durham, UK}

\begin{document}
\maketitle

\begin{abstract}
The problem of the contribution of cosmic rays to climate change is a
continuing one and one of importance. In principle, at least, the
recent results from the CLOUD project at CERN provide information
about the role of ionizing particles in 'sensitizing' atmospheric
aerosols which might, later, give rise to cloud droplets. Our analysis
shows that, although important in cloud physics the results do not
lead to the conclusion that cosmic rays affect atmospheric clouds
significantly, at least if $H_2SO_4$ is the dominant source of
aerosols in the atmosphere. An analysis of the 
very recent studies of stratospheric aerosol changes following a giant Solar Energetic 
Particles event shows a similar negligible effect. Recent measurements of the cosmic ray
 intensity show that a former decrease with time has been reversed. Thus, even if 
cosmic rays enhanced cloud production, there would be a small Global Cooling, not 
Warming.
\end{abstract}

\section{Introduction}
There is, by now, a wealth of literature on the relevance of cosmic rays (CR) to 
climate change. Examples favouring a significant effect are: Svensmark and 
Friis-Christensen (1997), Palle Bago and Butler (2000) and Svensmark (2007). Those 
against include Sloan and Wolfendale (2008) and  Erlykin et al. (2009 a,b).
The essence of the claim is that CR ions cause the nucleation of aerosols from trace 
condensable vapours, leading, via the growth of the aerosols, to cloud condensation 
nucleus (CCN) status and thereby to the water droplet stage. A separate aspect concerns
 the time variability of the CR intensity: recent work by 'CLOUD' 
(Kirkby et al., 2011) relating to laboratory experiments on the initial stages of cloud
condensation nuclei (CCN) have relevance to the claim and stratospheric nuclei 
(Mironova et al., 2012) caused by CR in the stratosphere are potentially important. 
Both studies will be examined here. 

In CLOUD, the nuclei are studied in a large `chamber' and ionizing particles come
 from an accelerator under carefully controlled conditions. Sulfuric acid vapour is 
studied in detail in that this is considered by Kirkby et al to be 'the primary vapour 
responsible for atmospheric nucleation'.
\section{Aerosols}
The mechanism by which cloud droplets form is one of some complexity and firstly we 
examine the process. Following Carslaw et al. (2002) the following stages can be 
identified. \\ 
1. For H$_2$O and H$_2$SO$_4$ vapours there can be condensation (~and evaporation~) 
leading to cluster nucleation, (ultra-fine condensation nuclei, UCN), the result being 
'subcritical embryos' of $<$1-2 nm diameter. \\
2. Condensation can occur yielding condensation nuclei (CN) - 'critical embryos' of 
diameter $\sim$1-2 nm. \\
3. Condensation and coagulation can then yield charged condensation nuclei of 
diameter $\sim$100 nm. \\
4. Activation follows and cloud droplets (CD), of diameter 10-20 $\mu$m, may then 
appear.

It will be apparent that different experiments will be responsive to different steps in
 the above: UCN, CN, CCN and CD. It is also clear that many (~often the majority~) of 
the UCN will not survive to CD.

In the CLOUD experiment, we assume that we are dealing largely with UCN and CN.

The stratospheric results relate to CCN, aerosols which are recognised by their 
extinction at wavelengths of 756 nm (~satellite studies~), and by other wavelengths 
(~eg in some cases at 360 nm~).

It is usually assumed (~eg Kirkby et al, 2011~) that sulfuric acid and ammonia are most
 relevant in the atmosphere and their effects have been examined in the CLOUD 
experiment, to be described next. 
\section{CLOUD.}
The main results from CLOUD, of relevance here, are:
\begin{itemize}
\item[(i)]`Atmospherically relevant ammonia' mixing ratios of 100 parts per trillion by
 volume increase the nucleation rate of sulphuric acid particles more than 
100-1000-fold. Of main importance to the present work, ions increase the nucleation 
rate by a factor of between 2 and 10 for ground level Galactic CR intensities.
\item[(ii)] The ion-induced nucleation can occur in the mid-troposphere but is 
negligible in the `boundary layer', ie below about 3km altitude. (~a fact remarked on 
by the CLOUD authors~).

Specifically, using the temperature dependence of the nucleation rate, CR - will only
 be relevant for Polar altitudes above about 4 km and Equatorial altitudes above about 
8 km, using the universally available temperature, altitude, latitude data. Thus, the 
boundary layer will be unaffected by CR-induced 
aerosols, at least for those involved in the CLOUD project, which are thought to be the
 ones of major importance (~an assumption that needs further analysis~).
\item[(iii)]There is a dramatic increase in nucleation rate with falling temperature 
(typically a factor $10^4$ in going from 292$^{\circ}$K to 248$^{\circ}$K), a result 
due to the change of saturation vapour pressure with temperature.
\end{itemize}
\section{Discussion of the CLOUD Aerosol Results.}
\subsection{General Remarks.}
Although aerosols are without doubt involved in the generation of CCN, their very 
uneven distribution across the Globe, for example, NOAA aerosol maps, (NOAA, 2012), in 
comparison with 
cloud cover, means that their effect is not straightforward. Their altitude dependence 
is a matter of importance and this will be examined. Early work by Elterman (1968) 
showed a slow fall in aerosol density with increaing height: at low altitudes the mean
attenuation coefficient was found to be about $3\cdot 10^{-2}km^{-1}$. More recently 
Hervig and Deshler (2002) have 
analysed comprehensive satellite- and balloon-borne data from the standpoint of the 
altitude variation of aerosols in 4 size ranges, from 386-1020 nm. The data relate to 
profiles over Laramie, Wyoming for 1984 to 1999. It is found that the density of 
aerosols is about 
constant with height from 12 km (~their lowest value~) to 20 km, after which it falls 
by a factor of 100 at 30 km. The relevance of this result will become clear later.  

Of interest to Cloud Cover is the reported slow increase of atmospheric aerosol density
 with time (4\% - 8\% y$^{-1}$ in the mid-latitude lower stratosphere, eg Liu et al., 
2012).
\subsection{Temperature dependence of nucleation.}
In the lower troposphere (the `boundary layer') the temperature is too high for 
nucleation to have any relevance to the CR, CC problem as already remarked. Since 
this is the region (LCC: `low cloud cover') where Svensmark and
 Friis-Christensen (1997) and others (~eg Palle Bago and Butler, 2000~), found the only
 evidence for CC,CR correlation then there is clearly no support for the CR, 
CC hypothesis. In the upper troposphere, where the temperature is lower - typically, at
 7.5km, the mean height of the high cloud cover (HCC) band, with $<$T$> \simeq$ 235K, 
the nucleation rate 
will be very high (from the CLOUD results) and a big CR, CC correlation would be 
expected. However, analyses such as our own, Erlykin et al. (2009a) shows no 
correlation at all for the HCC; this is despite the magnitude of the CR intensity (and 
its variation) being higher there, as well as the nucleation rate being predicted to be
 so high.

The dramatic temperature dependance of the nucleation rate would have other detectable 
effects on the measured CR, CC correlation even if present at altitudes where the 
temperatures are low enough for the predicted nucleation to be significant. 
Thus, there would be a strong latitude dependence due to the mean atmospheric 
temperature being a function of latitude; for example, at an altitude of 6km, $<$T$> 
\sim$ 263$^{\circ}$K at the Equator, 243$^{\circ}$K at latitude 45$^{\circ}$ and 220
$^{\circ}$K at latitude 80$^{\circ}$. Taken at face value the CLOUD results would 
indicate an increase in nucleation rate of about 3 orders of magnitude in going from 
the Equator to a latitude of 80$^{\circ}$. Even allowing for various reductions due to 
`sinks' (Kirkby, 2012, private communication) a big change should surely follow.

A search for the latitude dependence of the CR, LCC correlation, or the related 
dependence on the CR vertical rigidity cut-off (VRCO), gave negative results (Sloan and
 Wolfendale, 2008) and, indeed, this was one of the first demonstrations of the lack of
 a genuine CR, (L)CC correlation. A latitude dependence of the correlation was not 
detected at any altitude, in fact. Thus, the expected big change with latitude for 
H$_2$SO$_4$ nucleation anywhere is not observed.
\section{Aerosols in the Stratosphere}
\subsection{Ozone losses}
If CR are going to have an atmospheric effect anywhere it is surely in the 
stratosphere, where the CR intensities are so much higher than at ground level. This is
 particularly so for Solar Energetic Particles, 'SEP', events, where many of the
 protons (and heavier nuclei) lose all their energy in the stratosphere.

The influence of SEP on the ozone layer has been a topical subject for some years. 
Jackman et al. 
(1999) and others studied the effect of large SEP for the period 1965 to 1995 with 
positive results. They showed that the very large events of August 1972 and October 
1989 caused large increases in long-lived NO$_x$ constituents and that they caused 
direct ozone losses. Complications included differences between the seasons for impacts
 on polar ozone. 

Relevant work has been reported by Lu (2009). This worker claimed a correlation between
 CR and the polar ozone loss (~hole~) over Antarctica for the period 1980 - 2007 and 
predicted a severe loss in 2008-9. In fact, this loss was not observed. 

Our own 
analysis of yearly sunspot numbers and the area of the Antarctic ozone hole (~ the data
 coming from the NOAA, National Weather Service, 2012~) shows no correlation between 
them for the 
period for which there is accurate data: 1992 - 2011. Specifically, the correlation 
coefficient is 0.022, with a chance probability of a random association of 0.925. 
However, the above does not mean that there is no ozone-effect at all; for example,
Seppala et al. (2009) claim that large SEP events cause changes in polar surface 
temperatures. There is, though, the standard problem of distinguishing between CR- and 
Solar Intensity-induced effects; Lockwood (2012) makes the case for the latter.   
\subsection{Stratospheric aerosols.}
Of greater relevance in the present work is the effects of SEPs on aerosols in the 
stratosphere. Here, the work of Mironova et al. (2012) is important.

These workers concentrated on the effect on polar stratospheric aerosols of the extreme
 SEP event of 20 January 2005. They found evidence for the production of a 3-day 
'burst' of new particles 
and for the growth of pre-existing ultra-fine particles in the height range 10-20 km. 
Presumably these aerosols are in the size region exceeding 100 nm by their diameter. 
The evidence was only 
for limited longitude ranges (mainly in the N hemisphere, for the latitude studied, 
which was 66$^{\circ}$-73$^{\circ}$N).
    Our remarks on this interesting result are as follows.
\begin{itemize}
\item[(1)] The Polar Stratospheric Clouds (PSC), which may manifest the presence of new
 aerosols, only occur when the ambient temperature is low enough (~less than about
 200$^\circ$K~).
\item[(2)] Inspection of other data (~eg Mironova, 2011~) shows that the frequency of 
temperatures low enough to allow PSC formation (~by whatever mechanism~) has a dramatic
 time variation. Specifically, the PSC at Sodankyla, (~67$^\circ N, 27^\circ E$~) only 
occured in January and February, and occasionally in March. From 1965 to 2005 the 
January temperature was low enough for PSC for a percentage of the time varying from 
zero to 64\%. The value for January 2005 was 40\% and that for February 2005, the 
highest recorded in the 40 year period in question, was 80\%. The search for an 11-year
correlation of PSC is bedevilled by the concentration of PSC in the winter months, but 
for the 4 solar cycles reported by Mironova (2011) there is no evidence at all of 
higher rates for higher CR intensities and the equivalent for low CR intensities. PSC 
therefore require very special thermal condition.

As an upper limit for the generation of (new) clouds in the stratosphere we note that 
the actual percentages of the fraction of the 40-year period when such clouds can occur
 are 15\% for any cloud at all and 1\% for PSC to form with a 50\% probability.
\item[(3)] The chance probability of an increase in stratospheric extinction (quickly)
following a CR event (SEP) cannot yet be evaluated: only
one such coincidence has been detected. Even this event was complicated; the apparent 
SEP-initiated 'burst' of aerosols was followed by (~it is thought, unassociated~) new 
clouds; there was a burst of PSC some 5 days after the onset of the SEP. This burst of 
PSC was associated with a big reduction in temperature (~a fall of ~15$^\circ$C~) which
 was presumably of 'natural causes' - the influx of very cold air.
\item[(4)] Comparison of the mean longitudinal profile of the extinction coefficient 
for the 2005 event with those available for November 1978 to January 2009 (~McCormick 
et al., 1982~) - another period of very low atmospheric temperature - is relevant. 
Surprisingly, the 2005 event did not show a greater extinction magnitude at the 
greatest altitudes, such as would have been expected for CR-initiiation, the CR 
intensity increasing with altitude for the SEP event, but the aerosol density being 
constant (~see \S4.1~).
\end{itemize}
The implication of the forgoing is that even if the SEP event genuinely initiated a 
burst of aerosols, and that this burst could cause new clouds, then averaged 
over time (~see (2) above~) the effect of the stratospheric clouds will be very small 
and tropospheric implications largely absent. Thus, there is no evidence that SEP have 
a significant tropospheric climate effect, at least by way of CR ionization in the 
stratosphere.
\section{CR,Climate correlations in the lower atmosphere.}
That there might actually be a small, but spatially variable, effect of CR on Climate 
was shown by Voiculescu et al. (2006) who found evidence for a small
 CR, CC correlation for limited regions of the Globe from an analysis in which a 
distinction was made between CC changes initiated by solar UV and by CR. It was found 
that some 20\% - 30\% of the Earth's surface showed a negative correlation for the low 
cloud cover (LCC) and positive for the middle cloud cover (MCC). There is a 
complication, however, in that 
Erlykin et al. (2009b) claimed that both were due to convective flows of atmospheric 
air arising from changes in the solar irradiance and not due to CR at all.

It is relevant to point out that case for regional differences in climate change and 
correlations has been summarised by Lockwood (2012). This work related to solar-induced
 changes. Marked differences were found across the Globe but the integrated effect was
much less than the change due to anthropogenic sources.

Small effects of CR-induced droplet charging on cloud formation have been reported by
Harrison and Ambaum (2008).

\section{The Cosmic Ray Intensity.}
The early claims for a CR, CC correlation and its relevance to Global Warming relied on
 the CR intensity having fallen since records began: CR cause clouds and reduced clouds
 cause the warming. However, as we have shown elsewhere (Sloan and Wolfendale, 2011) 
the rate of CR fall was becoming smaller as the mean Global temperature was increasing 
rather rapidly. Furthermore, the neutron monitor (NM) data shows that the (smoothed) CR
 intensity `bottomed out' in the 1980s and has since increased. 

We have examined neutron monitor records of the ground level CR intensity from Oulu,
 Moscow and Jungfraujoch (~references given under names~). The values of the vertical 
rigidity cut-off ({\it VRCO}) are, respectively, 0.8, 2.3 and 4.5 GV. These values, 
particularly that for the Jungfraujoch can be regarded as representative for the Globe;
 this follows from the fact that for the region occupied largely by the Low Cloud 
Cover, $\langle VRCO \rangle \simeq$4.5 GV. Furthermore, for the region (~mainly 
Europe~) where Voiculescu et al. (2006) find evidence for a finite CR,LCC correlation, 
again $\langle VRCO \rangle \simeq$4.5 GV.
\begin{figure}[ht]
\begin{center}
\includegraphics[height=12cm,width=15cm]{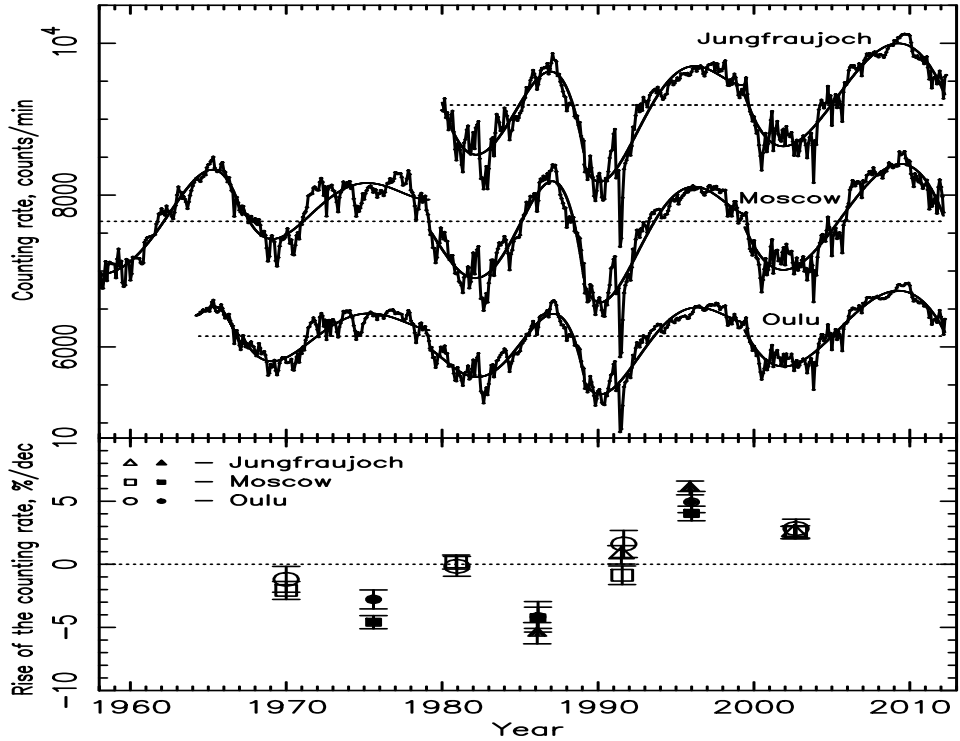}
\caption{\footnotesize The counting rate of Jungfraujoch (Jungfraujoch, 2012), Moscow 
(Moscow, 2012) and Oulu (Oulu, 2012) Neutron Monitors (upper panel), and the 
different rates of increase of the NM counts from peak to peak (~open symbols~) and 
minimum to minimum (~filled symbols~) (lower panel). Smooth full lines in the upper 
panel are 4-degree polynomial fits. The extent to which the smoothed CR intensity has 
been increasing for the last 20 years can be seen in the lower panel. The increase is 
marked; that for the higher, important, atmospheric levels, is higher (~see text~).}
\label{fig:Fig1}
\end{center}
\end{figure}

Figure 1 shows the time-series of the 3 sets of CR intensities. 4-degree polynomial 
fits have been made to each Solar Cycle, as shown. In order to examine the change of CR
intensity with time, in the presence of the 11-year Solar Cycle (~or, more accurately,
the rate of rise~) we examine the CR intensities (~NM rates~) at the same point on each
 Solar Cycle, specifically, the maximum and the minimum. It is apparent that each of 
the 7 datum years the data are reasonably consistent and no $VRCO$-dependence is 
apparent.

Further inspection shows that the rates of increase of the maximum NM rates are 
increasing linearly for the whole period. Those for the minima have fallen slightly 
before increasing rapidly. Figure 1 (upper panel) shows that the peak NM rates (~CR 
intensity~) have been increasing since about 1990 and the lower panel gives some 
evidence that there has been a steady upward change of the rate of increase of the
 maximum intensity and an irregular, but latterly high rate of increase of the CR 
intensity minima. Taken overall, from 1970 onwards, the rate of increase has been 
$2.6\pm 0.6$\% per decade (~a correlation coefficient of 0.69 and chance probability
0.087~).

The interpretation is interesting in its own right but mainly beyond the scope of the 
present work. Suffice it to say that there is a wealth of evidence pointing to 
important changes in solar properties in the period 1980 - 1990. Thus, there was a rise
before, and rapid fall after in the Solar magnetic moment (~eg Obridko and Shelting, 
2009~). Karam (2003) summarised the somewhat earlier data for solar ion flux, flow 
speed and ion density (~all at 1AU~) and found the same result: an increase until the
1980s and a fall thereafter.

Finally, in this area, it should be pointed out that the increase in CR ionization in 
the atmosphere above ground level is higher. Using the data of Bazilevskaya et al. 
(2008),
from the CR peak in 1987 to that in 1997 the increase for Mirny (~0.1 GV~) is 
$14\pm 4$\% for 8.2km altitude and $18\pm 4$\% for 5 km altitude. For the next cycle, 
(~1997 to 2009~) the values are $21\pm 5$\% for 8.2km and $29\pm 6$\% for 5km. For
Murmansk (~cut-off rigidity 0.5 GV~) the corresponding values are $2\pm 1$\% and 
$11\pm 3$\% for 8.2km altitude and $5\pm 2$\% and $13\pm 3$\% for 5km. The increases 
for the CR minima, from 1990 to 2002, are $25\pm 5$\% (8.2km) and $19\pm 4$\% (5km) for
Mirny and $14\pm 3$\% (8.2km) and $11\pm 3$\% (5km) for Murmansk.

There is no doubt that the CR intensity has been increasing significantly since the 
1980s.

That an increase is not unreasonable can be seen from studies of past sunspot records, 
the CR intensity being modulated by solar phenomena related strongly to sunspots.
Inspection of past sunspot records back to the commencement of telescopic observations 
in the early 1600s shows slow upward (~and downward~) movements of the smoothed SSN. 
Thus, slow changes in the (~strongly correlated~) CR intensity would occur.
Although prediction of the future trend in the CR intensity is hazardous, inspection of
 the sunspot data shows that after a peak the next higher SSN (~ie lower CR 
intensity~) can be 30-100 years away. Thus, it would not be surprising if the present 
rise in the smoothed CR intensity continued for several decades to come.
\section{Discussion and Conclusions.}
It is clear that the new results from CLOUD relating to aerosols indicate that although
 there could be a CR, Climate correlation by way of nucleation of aerosols, in the 
lower troposphere it should be very small indeed (a consequence of the `high' 
temperatures there). All pervading in the aerosol arguments, however, is the uncertain 
removal mechanisms which interpose themselves between the ultrafine condensation nuclei
 and embryos, all less than about 2nm in diameter (UCN,CN) and the condensation nuclei
 of 100nm and beyond (CCN,CD) stages (~see \S1~). In this context, Pierce and Adams 
(2009) quote loss factor (~for CN to CCN~) between 10 and 20 in the best case. New 
CLOUD studies are relating to the effect of aerosols other than those of $H_2SO_4$ and
ammonia which may be important in the atmosphere.

The fact that the CR intensity is rising again strongly militates further against a 
CR/Global Warming connection (in the absence of unphysically long phase lags).

There are other arguments against a CR/Climate correlation, not referred to above. 
These include a lack of atmospheric changes following nuclear explosions, nuclear 
accidents and natural radon variations (Erlykin et al., 2009a) and the non-observation 
of correlations for the type of cloud that should be responsive (if anything is) to CR 
ionization changes (Erlykin et al., 2009c).

{\Large {\bf Acknowledgements}} \\
We acknowledge the NMDB database {\it (www.nmdb.eu)}, founded under the European Union
FP7 programme (~contract no. 213007~) for providing data and also the staff of IGY JUNG
 monitor.

The authors are grateful to Professors Giles Harrison, Jasper Kirkby, Irina Mironova, 
Ilya Usoskin and Dr. Rolf Buetikofer for helpful correspondence.

We are grateful to the John C. Taylor Charitable Foundation for financial support.

{\Large {\bf References}} \\
Arnold F (2008) {\it Atmospheric ions and aerosol formation}. Space Sci. Rev. 
137, 225-239 \\
Bazilevskaya GA, Usoskin IG, Fl\"{u}ckiger EO et al. (2008) {\it Cosmic ray 
induced ion production in the atmosphere}. Space Sci. Rev. 137, 149-173 \\
Carslaw KS, Harrison RG, Kirkby J. (2002) {\it Cosmic rays, clouds and climate}. 
 Science 298, 1732-1737 \\
Elterman L, 1968, Air Force Cambridge Labs., AFCRL, 68-0153 \\
Erlykin AD, Gyalai G, Kudela K et al. (2009a) {\it Some aspects of ionization and
 cloud cover, cosmic ray correlation problem}. J.Atmos. and Solar-Terr. Phys. 71, 
823-829 \\
Erlykin AD, Gyalai G, Kudela K et al. (2009b) {\it On the correlation between 
cosmic ray intensity and cloud cover}. J.Atmos. and Solar-Terr. Phys. 71, 1794-1806 \\
Erlykin AD, Sloan T, Wolfendale AW (2009c) {\it The search for cosmic ray effects
 on clouds}. J.Atmos. and Solar-Terr. Phys. 71, 955-958 \\
Harrison RG, Ambaum MHP (2008), {\it Enhancement of cloud formation by droplet 
charging}. Proc. Roy. Soc. A, 464, 2561-2573 \\
Hervig M, Deshler T (2002) {\it Evaluation of aerosol measurements from SAGE II, 
HALOE and balloonborne optical particle counters}. J. Geophys. Res. 107, 4031-4042 \\
Jackman CH, Fleming EL, Vitt FM, Considine DB (1999) {\it The influence of solar 
proton events on the ozone layer}. Adv. Space Res. 24, 625-630 \\
Jungfraujoch (2012): {\it http://www.nmdb.eu/nest/search.php} \\
Karam PA (2003) {\it Inconstant Sun: how solar evolution has affected cosmic and 
ultraviolet radiation exposure over the history of life on Earth}. Health Physics 84, 
322-333 \\
Kirkby J, Curtius j, Almeida J et al. (2011) {\it Role of sulfuric acid, ammonia 
and galactic cosmic rays in atmospheric aerosol nucleation}.  Nature 476, 429-433 \\
Liu Y, Zhao X, Li W, Zhou X (2012) {\it Background stratospheric aerosol 
variations deduced from satellite observations}. J. Appl. Meteor. Climatol. 51, 799-812  \\
Lockwood M. (2012) {\it Solar influence on global and regional climates}. Surveys
 in Geophys. 33, 503-534 \\
Lu Q-B. (2009) {\it Correlation between cosmic rays and ozone depletion}. Phys. 
Rev. Lett. 102, 118501(4pp) \\
McCormick MP, Steele HM, Hamill P et al. (1982), {\it Polar stratospheric cloud 
sightings by SAM II}. J. Atmos. Sci. 39, 1387-1397 \\  
Mironova IA (2011), COST Action ES1005 \\
Mironova IA, Usoskin IG, Kovaltsov GA, Petelina SV (2012), {\it Possible effect 
of extreme solar energetic particle event of 20 January 2005 on polar stratospheric 
aerosols: direct observational evidence}. Atmos. Chem. Phys. 12, 769-778 \\
Moscow (2012): {\it http://helios.izmiran.rssi.ru/cosray/main.htm} \\
NOAA National Weather Service (2012) Climate Prediction Centre -  \\
aerosol optical thickness: {\it http://www.nws.noaa.gov} \\
Obridko VN, Shelting BD (2009) {\it Anomalies in evolution of global and large-
scale solar magnetic fields as the precursor of several upcoming low solar cycles}. 
Astron. Lett. 35, 247-252 \\
Oulu (2012): {\it http://cr0.izmiran.rssi.ru/oulu/main.htm} \\
Palle Bago E, Butler CJ (2000) {\it The influence of cosmic rays on terrestrial 
clouds and global warming}. Astron. Geophys. 41, 4.18-4.22 \\
Pierce JR, Adams PJ (2009) {\it Uncertainty in global CCN concentrations from 
uncertain aerosol nucleation and primary emission rates}. Atmos. Chem. Phys. 9, 
1339-1356 \\
Sepp\"{a}l\"{a} A, Randall CE, Clilverd MA et al. (2009) {\it Geomagnetic 
activity and polar surface air temperature variability}, J. Geophys. Res. 114, 
A10312(10pp) \\
Sloan T, Wolfendale AW (2008) {\it Testing the proposed causal link between 
cosmic rays and cloud cover}. Env. Res. Lett. 3, 024001(6pp) \\
Sloan T, Wolfendale AW (2011) {\it The contribution of cosmic rays to global 
warming}. J. Atmos. Solar-Terr. Phys. 73, 2352-2355 \\
Svensmark H, Friis-Christensen E (1997) {\it Variation of cosmic ray flux and 
global cloud coverage - a missing link in solar-climate relationship}. J. Atmos. Sol. 
Terr. Phys. 59, 1225-1232 \\
Svensmark H (2007) {\it Cosmoclimatology: a new theory emerges}. News and Rev. in
 Astron. Geophys. 48, 1.18-1.24 \\
Voiculescu M, Usoskin IG, Mursula K  (2006) {\it Different response of clouds to 
solar input}. Geophys. Res. Lett., 33, L21802(5pp) 
\end{document}